\documentclass[final]{IEEEtran}
\newlength \figwidth
\usepackage{cite}
\ifCLASSINFOpdf
  \usepackage[pdftex]{graphicx}
	\usepackage{epstopdf}
  \graphicspath{{../Figures/}}
  \DeclareGraphicsExtensions{.eps}
\else
  \usepackage[dvips,draft]{graphicx}
\fi
\usepackage[cmex10]{amsmath}
\usepackage{amssymb}
\usepackage{pifont}
\usepackage{amsthm}
\usepackage{url}
\usepackage{dsfont}
\usepackage{tabulary}
\usepackage{multirow}
\usepackage{color}
\newcounter{MYtempeqncnt}

\usepackage{color}
\usepackage{times}
\usepackage{verbatim}
\usepackage{floatflt}
\usepackage{enumerate}
\usepackage{array}
\usepackage{multicol,afterpage,wrapfig} 
\usepackage{tikz}
\usepackage[noend]{algpseudocode}
\makeatletter
\def\BState{\State\hskip-\ALG@thistlm}
\makeatother
\usepackage{amsmath,amssymb,eucal}
\usepackage[utf8]{inputenc}
\usepackage{epsfig}
\usepackage{exscale}
\usepackage{latexsym}
\usepackage{verbatim}
\usepackage{amsfonts}
\usepackage{multirow}
\usepackage{cite}
\usepackage{balance}
\usepackage{color}
\usepackage{transparent}
\usepackage{import}
\usepackage{mathtools}
\usepackage{tikz}
\usetikzlibrary{automata,arrows,positioning,calc}
\usepackage{bbm}
\usepackage[printonlyused,withpage]{acronym}
\usepackage{tabu}
\usepackage{longtable}
\usepackage{balance}
\usepackage{footnote}
\usepackage{tablefootnote}
\usepackage{adjustbox}
\usepackage{multirow}
\usepackage{pifont}
\def\BibTeX{{\rm B\kern-.05em{\sc i\kern-.025em b}\kern-.08em
    T\kern-.1667em\lower.7ex\hbox{E}\kern-.125emX}}
    
\renewcommand{\IEEEQED}{\IEEEQEDopen}

\newcommand*\xbar[1]{%
  \hbox{%
    \vbox{%
      \hrule height 0.5pt 
      \kern0.36ex
      \hbox{%
        \kern-0.12em
        \ensuremath{#1}%
        \kern-0.12em
      }%
    }%
  }%
}

\newfont{\bbb}{msbm10 scaled 500}

\newfont{\bb}{msbm10 scaled 1100}

\newcommand{\executeiffilenewer}[3]{%
\ifnum\pdfstrcmp{\pdffilemoddate{#1}}%
{\pdffilemoddate{#2}}>0%
{\immediate\write18{#3}}\fi%
}

\usetikzlibrary{arrows,calc}
\allowdisplaybreaks

\IEEEoverridecommandlockouts
\begin{document}
\pagenumbering{gobble}

\newtheorem{Theorem}{\bf Theorem}
\newtheorem{Corollary}{\bf Corollary}
\newtheorem{Remark}{\bf Remark}
\newtheorem{Lemma}{\bf Lemma}
\newtheorem{Proposition}{\bf Proposition}
\newtheorem{Assumption}{\bf Assumption}
\newtheorem{Definition}{\bf Definition}
\title{Supporting UAV Cellular Communications\\through Massive MIMO}
\author{\IEEEauthorblockN{{Giovanni~Geraci$^{\star}$, Adrian~Garcia-Rodriguez$^{\star}$, Lorenzo~Galati~Giordano$^{\star}$, David~L\'{o}pez-P\'{e}rez$^{\star}$, and Emil Bj\"{o}rnson$^{\dagger}$}}\\ \vspace{0.2cm}
\IEEEauthorblockA{$^{\star}$\emph{Nokia Bell Labs, Dublin, Ireland}\\
\IEEEauthorblockA{$^{\dagger}$\emph{Department of Electrical Engineering (ISY), Link\"{o}ping University, Link\"{o}ping, Sweden}
}}}
\maketitle
\thispagestyle{empty}
\begin{abstract}
In this article, we provide a much-needed study of UAV cellular communications, focusing on the rates achievable for the UAV downlink command and control (C\&C) channel. For this key performance indicator, we perform a realistic comparison between existing deployments operating in single-user mode and next-generation multi-user massive MIMO systems. We find that in single-user deployments under heavy data traffic, UAVs flying at 50\,m, 150\,m, and 300\,m achieve the C\&C target rate of 100~kbps -- as set by the 3GPP -- in a mere 35\%, 2\%, and 1\% of the cases, respectively. Owing to mitigated interference, a stronger carrier signal, and a spatial multiplexing gain, massive MIMO time division duplex systems can dramatically increase such probability. Indeed, we show that for UAV heights up to 300\,m the target rate is met with massive MIMO in 74\% and 96\% of the cases with and without uplink pilot reuse for channel state information (CSI) acquisition, respectively. On the other hand, the presence of UAVs can significantly degrade the performance of ground users, whose pilot signals are vulnerable to UAV-generated contamination and require protection through uplink power control.
\end{abstract}
\IEEEpeerreviewmaketitle

\section{Introduction}

Whether we greet their latest proliferation with thrill or dismay, unmanned aerial vehicles (UAVs) will likely be the best candidates to automate and ease many of our critical tasks in the near future. Example of these include emergency search and rescue after natural disasters, such as earthquakes or flash floods, crowd management, and surveillance for public safety. Thanks to their reduced cost, UAVs will also be suitable for less vital applications such as parcel delivery and video streaming of spectacular landscapes. All but unheard of until just recently, UAVs now promise to play a key technological and commercial role for years to come \cite{NewAmerica2015,ZenZhaLim2016}. 
 
For such optimistic vision to come true, UAVs will necessitate control and connectivity over a wireless network. Notably, reliable command and control (C\&C) channels to the UAVs are required to safely operate these vehicles remotely and beyond current visual line-of-sight (LoS) constraints. In this setup, terrestrial cellular networks are well positioned to serve UAVs flying up to an altitude of few hundred meters by partially reusing the existing network infrastructure and spectrum resources while undergoing the necessary upgrades \cite{HayYanMuz2016,BerChiPol2016}. With the aim of developing an enhanced cellular support for UAV communications, the Third Generation Partnership Project (3GPP) has been gathering key industrial players, producing systematic measurements and accurate modeling of UAV-to-ground channels \cite{3GPP36777}. The remarkable industrial involvement in this standardization fora \cite{RP170779,R11717287,R11718267,R11714071}, together with the concurrent theoretical investigations being undertaken in academia (see, e.g., \cite{WanJiaHan2017,ChaDanLar2017,AzaRosPol2017, GalKibSilva2017, MozSaaBen2017} and references therein), motivate our effort in bridging the gap between present- and forward-looking aerial communications research.

In this article, we aim at advancing the understanding of UAV cellular communications, paying particular attention to the performance of the UAV downlink (DL) C\&C channel, for which a minimum requirement of 100~kbps has been defined \cite{3GPP36777}. We study two network architectures: \emph{(i)} a traditional network with sectorized BSs operating in \emph{single-user} mode -- representing most existing cellular deployments; and \emph{(ii)} a massive MIMO cellular network operating in \emph{multi-user} mode with digital beamforming capabilities -- exemplifying next-generation deployments. For these practical scenarios, we examine how a UAV's height affects its cell selection process and its performance. We also evaluate the increased reliability that can be achieved for the UAV C\&C channel through massive MIMO, and we quantify what the presence of UAVs entails for the performance of conventional ground UEs (GUEs). The main takeaways of this paper can be summarized as follows:
\begin{itemize}
\item In single-user mode deployments, UAVs taking off or landing at 1.5\,m achieve the C\&C target rate of 100~kbps in 87\% of the cases. However, because of strong LoS interference received from a plurality of visible cells, such reliability decreases to 35\% when they fly at 50\,m, and to a mere 2\% and 1\% at 150\,m and 300\,m, respectively.
\item Multi-user massive MIMO systems can support a 100~kbps C\&C channel for UAV heights up to 300\,m with 74\% and 96\% reliability with and without pilot reuse, respectively. This is due to mitigated interference, a stronger carrier signal, and a spatial multiplexing gain.
\item The presence of UAVs can significantly degrade the performance attained by GUEs with massive MIMO. Uplink (UL) power control policies are required to protect GUEs, whose pilot signals are otherwise vulnerable to severe contamination from UAV-generated overlapping pilots.
\end{itemize}

\begin{figure*}[!t]
\centering
\includegraphics[width=\figwidth]{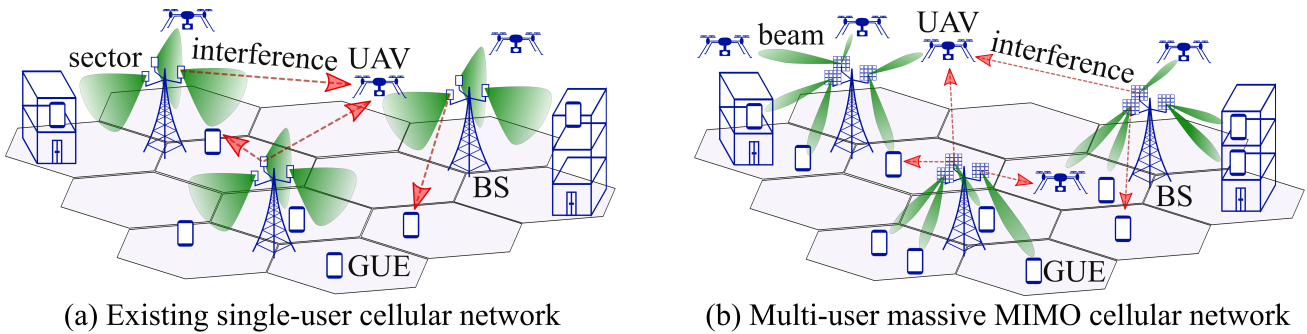}
\caption{Illustration of two examples of cellular infrastructure to support both ground and UAV cellular communications. In (a) -- similarly to most existing networks -- BSs are equipped with a vertical antenna panel to cover a cellular sector and serve a single user on each PRB, potentially generating strong interference at neighboring users. In (b) -- exemplifying next-generation networks -- BSs are equipped with massive MIMO arrays and serve multiple users on each PRB through beamforming, also increasing the useful signal at each served user and mitigating the interference at neighboring users.}
\label{fig:Network}
\end{figure*}
\section{3GPP System Model}
\label{sec:System_Model}

In this section, we introduce the 3GPP network topology and channel model used in this paper. Further details on the specific parameters used for our studies are summarized in Table~\ref{table:parameters}.

\subsection{Cellular Network Deployment}

In this paper, we concentrate on the DL of a time division duplex (TDD) cellular network as illustrated in Fig.~\ref{fig:Network}, where BSs are deployed on a hexagonal layout to communicate with their associated users. A deployment site is comprised of three co-located BSs, each providing coverage to one sector of $120^{\circ}$. Unlike conventional cellular networks, the network under consideration serves both GUEs and UAVs, e.g., providing the former with DL data streams and the latter with C\&C information. In the following, we let the term \emph{users} denote both GUEs and UAVs. As specified by the 3GPP in \cite{3GPP36777, 3GPP38901}, GUEs are deployed both outdoor (at a height of 1.5\,m) and indoor in buildings comprised of $4$ to $8$ floors. Instead, UAVs are deployed outdoor at variable heights between 1.5\,m, which characterizes their height during take off and landing, and 300\,m, which represents their maximum cruising altitude with cellular service \cite{3GPP36777}.

We denote by $\mathcal{B}$  the set of cellular BSs, and assume that all BSs are equipped with $N_{\mathrm{a}}$ antennas and transmit with power $P_{\textrm{b}}$. Users associate to the BS that provides the largest average reference signal received power (RSRP) and are equipped with a single antenna. Therefore, the total number of associated users per BS is determined by their physical location and their propagation characteristics towards their neighboring BSs. We denote as $\mathcal{K}_b$ the set of users served by BS $b$ on a given physical resource block (PRB), and by $K_b$ its cardinality. We remark that the set $\mathcal{K}_b$ can be dynamically defined by the $b$-th BS through scheduling operations.

\subsection{3D Propagation Channel}

In this paper, we adopt the newly released 3GPP channel model designed for evaluating the performance of cellular networks with UAVs \cite{3GPP36777}. Among other real-world propagation phenomena, this model accounts for 3D channel directionality, antenna polarization, spatially correlated large-scale parameters such as shadow fading, and time-and-frequency correlated fast fading. All the parameters in the model have been derived from measurement campaigns carried out to characterize the particular propagation features of both GUEs and UAVs, and therefore they explicitly account for the transmitter and receiver heights  \cite{3GPP36777, 3GPP38901}. 

Let $\mathbf{h}_{bjk} \in \mathbb{C}^{N_{\mathrm{a}}\times 1}$ represent the channel between BS $b$ and user $k$ in cell $j$ on a specific PRB. The signal $y_{bk} \in \mathbb{C}$ received by user $k$ in cell $b$ can be expressed as (\ref{eqn:rx_signal}), where $s_{bk} \in \mathbb{C}$ is the unit-variance signal intended for user $k$ in cell $b$, $\epsilon_{bk} \sim \mathcal{CN}(0,\sigma^2_{\epsilon})$ is the thermal noise, and $\mathbf{w}_{bk} \in \mathbb{C}^{N_{\mathrm{a}}\times 1}$ is the spatial precoder used by BS $b$ to serve user $k$ in cell $b$. The four terms on the right hand side of (\ref{eqn:rx_signal}) respectively represent: useful signal, intra-cell interference from the serving BS, inter-cell interference from other BSs, and thermal noise.

The signal-to-interference-plus-noise ratio (SINR) $\gamma_{bk}$ experienced by the $k$-th user associated to the  $b$-th BS on a given PRB is obtained via an expectation over all symbols, and it is given by (\ref{equ:sinr_ue}). To accurately characterize the performance of the considered setup, we map each SINR value to the rate delivered on a given PRB by considering link adaptation, i.e., selecting the maximum modulation and coding scheme (MCS) that ensures a desired block error rate (BLER) \cite{3GPP36213}. We set the BLER to $10^{-1}$, which is a sufficiently low value considering that retransmissions further reduce the number of errors. This yields a minimum spectral efficiency of 0.22~b/s/Hz for SINRs between -5.02~dB and -4.12~dB, and a maximum spectral efficiency of 7.44~b/s/Hz for SINRs of 25.87~dB and above. Moreover, we also account for the overhead introduced by control signaling when computing the resultant user rates \cite{3GPP36213}.

\begin{table}
\centering
\caption{System parameters}
\label{table:parameters}
\def\arraystretch{1.2}
\begin{tabulary}{\columnwidth}{ |p{2.6cm} | p{5.4cm} | }
\hline
	BS distribution				& Three-tier wrapped-around hexagonal grid, 37 sites, three sectors each, one BS per sector \cite{3GPP36777} \\ \hline
  BS inter-site distance 		& 500\,m \cite{3GPP36777} \\ \hline
  User distribution 				& 15 users per sector on average \cite{3GPP36777} \\ \hline
	\multirow{2}{*}{GUE distribution} 				& 80\% indoor, horizontal: uniform, vertical: uniform in buildings of four to eight floors  \\ \cline{2-2}
	 				& 20\% outdoor, horizontal: uniform, vertical: 1.5\,m \\ \hline
	UAV distribution 				& 100\% outdoor, horizontal: uniform, vertical: uniform between 1.5\,m and 300\,m \cite{3GPP36777} \\ \hline
	UAVs/GUEs ratio 				& 3GPP Case~3: 7.1\% \cite{3GPP36777} \\ \hline
	User association				& Based on RSRP (slow channel gain) \\ \hline
	Path loss, prob. LoS, shadowing, fast fading 					& Urban Macro as per \cite{3GPP36777}  \\ \hline 
	\multirow{2}{*}{Channel estimation}		& Single-user: perfect channel estimation\\ \cline{2-2}
			& Multi-user: UL SRSs with Reuse 3 \\ \hline
	Control overhead 		& 3 OFDM symbols per PRB \cite{3GPP36213}\\ \hline 
	Thermal noise 				& -174 dBm/Hz spectral density \cite{3GPP36777}\\ \hline 
	Noise figure 			& BS: 7~dB, user: 9~dB \cite{3GPP36777} \\ \hline
	Carrier frequency 		& 2~GHz \cite{3GPP36777} \\ \hline
	System bandwidth 		& 10 MHz with 50 PRBs \cite{3GPP36777} \\ \hline
	BS transmit power 	     & 46~dBm \cite{3GPP36777} \\ \hline   
	BS antenna elements 		& Horizontal and vertical half power beamwidth: $65^{\circ}$, max. gain: 8~dBi \cite{3GPP36777} \\ \hline
	BS array configuration 		& Height: $25$m, mechanical downtilt: $12^{\circ}$, element spacing: $0.5\lambda$ \cite{3GPP36777}\\ \hline
	\multirow{2}{*}{BS array size } 		& Single-user: $8\times 1$ X-POL $\pm 45^{\circ}$, 1 RF chain \\ \cline{2-2}
	  		& Multi-user: $8\times 8$ X-POL $\pm 45^{\circ}$, 128 RF chains  \\ \hline
	\multirow{2}{*}{BS precoder} 		& Single-user: none \\ \cline{2-2}
	  		& Multi-user: zero-forcing  \\ \hline
  \multirow{2}{*}{Power control}		& DL: equal power allocation\\ \cline{2-2}
			& UL: fractional power control with $\alpha = 0.5$, $P_{0} = -58$~dBm, and $P_{\textrm{max}}=23$~dBm \cite{UbeVilRos2008}\\ \hline
	User antenna 		& Omnidirectional with vertical polarization, gain: 0~dBi \cite{3GPP36777} \\ \hline
	Traffic model		& Full buffer \\ \hline
	\multirow{2}{*}{Scheduler}		& Single-user: round-robin, one user per PRB \\ \cline{2-2}
			& Multi-user: round-robin, eight users per PRB \\ \hline
\end{tabulary}
\end{table}
\begin{figure*}[!t]
\normalsize
\setcounter{MYtempeqncnt}{\value{equation}}
\hrulefill
\begin{align}
y_{bk} = \!\sqrt{P_{\textrm{b}}} \, {\mathbf{h}_{bbk}^{\mathrm{H}} \mathbf{w}_{bk} s_{bk}} + \!\sqrt{P_{\textrm{b}}} \!\!\!\! \sum_{i\in\mathcal{K}_b\backslash k} \!\!\!\!\!  {\mathbf{h}_{bbk}^{\mathrm{H}} \mathbf{w}_{bi} s_{bi}}
+ \! \sqrt{P_{\textrm{b}}} \! \sum_{j \in \mathcal{B} \backslash b} \,\sum_{\,i\in\mathcal{K}_{j}} \!\! {\mathbf{h}_{jbk}^{\mathrm{H}} \mathbf{w}_{ji} s_{ji}} \!+\! \epsilon_{bk}
\label{eqn:rx_signal}
\end{align}
\setcounter{equation}{\value{MYtempeqncnt}}
\addtocounter{equation}{1}
\setcounter{MYtempeqncnt}{\value{equation}}
\begin{align}\label{equ:sinr_ue}
\gamma_{bk} = \frac{ P_{\textrm{b}} \, \vert \mathbf{h}_{bbk}^{\mathrm{H}} \mathbf{w}_{bk} \vert^2 }
{P_{\textrm{b}} \sum_{i\in\mathcal{K}_b\backslash k} \vert \mathbf{h}_{bbk}^{\mathrm{H}} \mathbf{w}_{bi} \vert^2 + 
P_{\textrm{b}} \sum_{j \in \mathcal{B} \backslash b} \sum_{i\in\mathcal{K}_{j}} \vert \mathbf{h}_{jbk}^{\mathrm{H}} \mathbf{w}_{ji} \vert^2 + \sigma^2_{\epsilon}}
\end{align}
\hrulefill
\setcounter{equation}{\value{MYtempeqncnt}}
\vspace*{-3pt}
\end{figure*}
\addtocounter{equation}{1}

\section{UAV performance in Single-User Mode}
\label{sec:SU-MIMO}

In this section, we consider a cellular network as depicted in Fig.~\ref{fig:Network}(a), where the BSs are equipped with $N_{\mathrm{a}}=16$ antennas arranged in a vertical array of 8 cross-polarized (X-POL) elements and connected to a single radio-frequency (RF) chain. Each BS schedules a maximum of one user on each PRB using fixed analog beamforming with a $12^{\circ}$ mechanical antenna downtilt. We denote this setup as a \emph{single-user} scenario, and we consider it to exemplify a large number of existing cellular deployments. In this single-user setting, equations (\ref{eqn:rx_signal}) and (\ref{equ:sinr_ue}) are simplified as follows: all vectors $\mathbf{w}$ are unit-norm and consist of identical scalars, the second term on the right hand side of (\ref{eqn:rx_signal}) vanishes and so does the first term in the denominator of (\ref{equ:sinr_ue}), and all double sums reduce to single sums. For this single-user scenario, we will first examine how UAVs associate to BSs depending on the height of the former, and we will then show how the performance of a UAV C\&C channel is affected by its height.

\subsection{UAV Cell Selection}

Fig.~\ref{fig:SU-AntennaGain} depicts the antenna gain between a BS and a UAV aligned to the BS's horizontal bearing as a function of the 2D ground distance between them. Said antenna gain is plotted for UAV heights of 1.5\,m, 50\,m, 75\,m, 150\,m, and 300\,m, allowing to identify different behaviors for different UAV height ranges. 

\subsubsection*{Low UAVs}
A UAV at 1.5\,m falls within the main lobe of a BS as long as its 2D distance exceeds 52m. Moreover, the antenna gain is maximized for 2D distances in the range 80\,m-180\,m. Considering that the ISD is 500\,m, the closest BS provides the maximum antenna gain, and any associations to BSs other than the closest one are mainly explained by differences in shadow fading and LoS conditions.

\subsubsection*{High UAVs}
\begin{figure}[!t]
\centering
\includegraphics[width=\columnwidth]{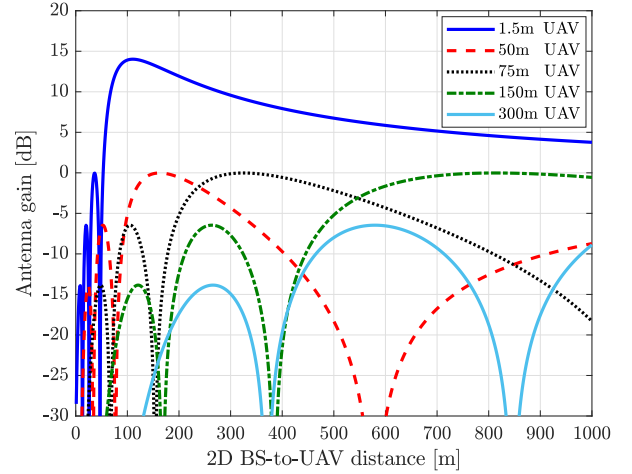}
\caption{Antenna gain between a BS and a UAV aligned to the BS's horizontal bearing as a function of the 2D distance between them and for various UAV heights.}
\label{fig:SU-AntennaGain}
\end{figure}

\begin{figure}[!t]
\centering
\includegraphics[width=\columnwidth]{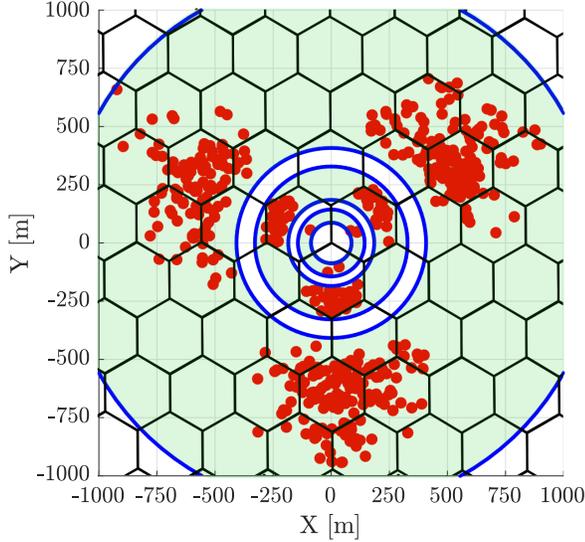}
\caption{2D location samples (red dots) of 150\,m-high UAVs associated to a three-sector BS deployment site located at the origin. Distance ranges corresponding to different secondary lobes are shaded and delimited by blue circles.}
\label{fig:SU-Association}
\end{figure}

As a UAV's height increases to 50\,m and beyond, the main BS antenna lobe is only visible at 2D distances larger than 1km (outside the range of Fig.~\ref{fig:SU-AntennaGain}). The secondary lobes thus play a significant role in the association process. Owing to the almost free-space propagation experienced by high-altitude UAVs, the path loss difference between the closest and the further away BSs is not significant compared to the difference in their respective antenna gains \cite{3GPP36777}. As a result, the UAV tends to associate to BSs located a few tiers away. This phenomenon is further illustrated in Fig.~\ref{fig:SU-Association}, which adopts the perspective of a three-sector BS located at the origin, and shows samples of the 2D locations of its associated UAVs (red dots) for UAV heights of 150\,m. Fig.~\ref{fig:SU-Association} confirms the existence of distance ranges (represented by the green shaded regions), each corresponding to one of the secondary lobes in Fig.~\ref{fig:SU-AntennaGain}.

\subsection{UAV C\&C Channel Performance}

\begin{figure}[!t]
\centering
\includegraphics[width=\columnwidth]{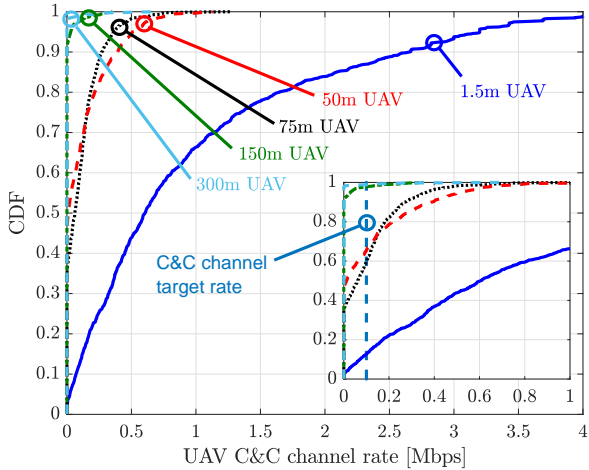}
\caption{CDF of the UAV C\&C channel rates as a function of the UAV height in a single-user scenario. In the enlargement, the target rate of 100~kbps is also shown for comparison.}
\label{fig:SU-McsRate_CDF}
\end{figure}

Fig.~\ref{fig:SU-McsRate_CDF} shows the data rate performance of the UAV C\&C channel over a 10~MHz bandwidth for various UAV heights, motivating the following observations:
\begin{itemize}
\item UAVs at a height of 1.5\,m achieve the target rate of 100~kbps 87\% of the time. Moreover, 34\% of the time their data rates even exceed 1~Mbps.
\item UAVs flying at around 50\,m and 75\,m only achieve the target rate 35\% and 40\% of the time, respectively. The achievable rates for this range of UAV heights almost never reach 1~Mbps (0.3\% of the time).
\item As UAVs fly higher, the target rate of 100~kbps can only be achieved for a small fraction of time, amounting to just 2\% and 1\% for heights of 150\,m and 300\,m, respectively.
\end{itemize}
The above results allow to conclude that cellular networks with heavy data traffic and that simply rely on single-user mode operations are unlikely to be able to support the much-needed C\&C channel for UAVs flying at reasonable heights.

\section{UAV Performance in Massive MIMO Systems}
\label{sec:MU-MIMO}

In this section, we consider a cellular network as depicted in Fig.~\ref{fig:Network}(b), where BSs are equipped with massive MIMO arrays and provided with digital beamforming and spatial multiplexing capabilities. In particular, we consider $N_{\mathrm{a}}=128$ antennas, arranged in an $8\times 8$ planar array of X-POL elements, supported by 128 RF chains. Since massive MIMO allows the transmission of beamformed control channels, users tend to associate to the closest BS. We allow each BS $b$ to serve a maximum of $K_b=8$ users per PRB through digital zero forcing (ZF) precoding. We denote this setup as a \emph{multi-user} scenario, and we consider it to exemplify next-generation massive MIMO cellular deployments.

\subsection{Preliminaries}

In this multi-user setting, the network operates in a TDD fashion, where channels are estimated at the BS through the use of UL sounding reference signals (SRSs) -- commonly known as \emph{pilots} -- sent by the users under the assumption of channel reciprocity \cite{MarLarYanBook2016,GalCamLopWCNC2017}.

\subsubsection*{Channel state information acquisition}
Let the pilot signals span $M_{\mathrm{p}}$ symbols. The pilot transmitted by user $k$ in cell $b$ is denoted by $\mathbf{v}_{\textrm{i}_{bk}} \in \mathbb{C}^{M_{\mathrm{p}}}$, where $\textrm{i}_{bk}$ is the index in the pilot codebook, and all pilots form an orthonormal basis \cite{MarLarYanBook2016}. Each pilot signal received at the BS suffers contamination due to pilot reuse across cells. We assume Reuse 3, i.e., each pilot signal is orthogonal among the three $120^{\circ}$ BS sectors of the same site, but it is reused among all BS sites, generating contamination. This solution is particularly practical from an implementation standpoint, since it involves coordination only between the three co-located BSs of the same BS deployment site. The collective received signal at BS $b$ is denoted as $\mathbf{Y}_b \in \mathbb{C}^{N_{\mathrm{a}} \times M_{\mathrm{p}}}$, and given by
\begin{equation}
\mathbf{Y}_b = \sum_{j \in \mathcal{B}} \sum_{k\in\mathcal{K}_j} \sqrt{P_{jk}} \mathbf{h}_{bjk} \mathbf{v}_{\mathrm{i}_{jk}}^{\textrm{T}} + \mathbf{N}_b,	
\label{eqn:Yb}
\end{equation}
where $\mathbf{N}_b$ contains the additive noise at BS $b$ during pilot signaling with independent and identically distributed entries following $\mathcal{CN}(0,\sigma^2_{\epsilon})$, and $P_{jk}$ is the power transmitted by user $k$ in cell $j$. We consider the following fractional UL power control rationale \cite{UbeVilRos2008}
\begin{equation}
P_{jk} = \min\left\{ P_{\textrm{max}}, P_0 \cdot \bar{h}_{jjk}^\alpha \right\},
\label{eqn:power_control}
\end{equation}
where $P_{\textrm{max}}$ is the maximum user transmit power, $P_0$ is a cell-specific parameter, $\alpha$ is a path loss compensation factor, and $\bar{h}_{jjk}$ is the slow channel gain measured at UE $k$ in cell $j$ based on the RSRP~\cite{3GPP36213}. The aim of (\ref{eqn:power_control}) is to compensate only for a fraction $\alpha$ of the path loss, up to a limit imposed by $P_{\textrm{max}}$.

The received signal $\mathbf{Y}_b$ in (\ref{eqn:Yb}) is processed at BS $b$ by correlating it with the known pilot signal $\mathbf{v}_{\textrm{i}_{bk}}$, thus rejecting interference from other orthogonal pilots. The $b$-th BS therefore obtains the following least-square channel estimate for user $k$ in cell $b$
\begin{equation}
\begin{aligned}
\widehat{\mathbf{h}}_{bbk} &=
\frac{1}{\sqrt{P_{bk}}} \mathbf{Y}_b \mathbf{v}_{\mathrm{i}_{bk}}^{*} = \mathbf{h}_{bbk} \\
& \enspace+ \frac{1}{\sqrt{P_{bk}}} \Big( \sum_{j \in \mathcal{B} \backslash b} \,\sum_{k\in\mathcal{K}_j} \sqrt{P_{jk}} \mathbf{h}_{ijk} \mathbf{v}_{\mathrm{i}_{jk}}^{\textrm{T}} +
\mathbf{N}_i \Big) \mathbf{v}_{\mathrm{i}_{bk}}^{*}
\label{eqn:PC}
\end{aligned}
\end{equation}
where intra-cell pilot contamination is not present since BS $b$ allocates different pilots for the users in its own cell.

\subsubsection*{Spatial multiplexing through digital precoding}
Each BS simultaneously serves multiple users on each PRB through ZF precoding, attempting to suppress all intra-cell interference. Let us define the estimated channel matrix $\widehat{\mathbf{H}}_b \in \mathbb{C}^{N_{\mathrm{a}}\times K_b}$ as
\begin{equation}
\widehat{\mathbf{H}}_b = \left[ {\widehat{\mathbf{h}}_{bb1}},\ldots,{\widehat{\mathbf{h}}_{bb{K_b}}} \right].
\label{eqn:Hb}
\end{equation}
The ZF precoder
\begin{equation}
	\mathbf{W}_b = \left[ {\mathbf{w}}_{b1},\ldots,{\mathbf{w}}_{b K_{b}} \right]	
\label{eqn:precoder}
\end{equation}
at BS $b$ can be calculated as \cite{SpeSwiHaa:04}
\begin{equation}
	\mathbf{W}_b = \widehat{\mathbf{H}}_b \left( \widehat{\mathbf{H}}_b^{\mathrm{H}} \widehat{\mathbf{H}}_b \right)^{-1} \left(\mathbf{D}_b\right)^{-\frac{1}{2}},
	\label{eqn:ZF}
\end{equation}
where the diagonal matrix $\mathbf{D}_b$ is chosen to meet the transmit power constraint with equal user power allocation, i.e., $\Vert \mathbf{w}_{bk}\Vert^2 = P_{\mathrm{b}}/K_b$ $\forall k, b$. The SINR on a given PRB for user $k$ can be calculated from (\ref{equ:sinr_ue}), with the precoding vectors $\mathbf{w}_{bk}$ obtained as in (\ref{eqn:precoder}).


\subsection{UAV C\&C Channel Performance Improvement}

\begin{figure}[!t]
\centering
\includegraphics[width=\columnwidth]{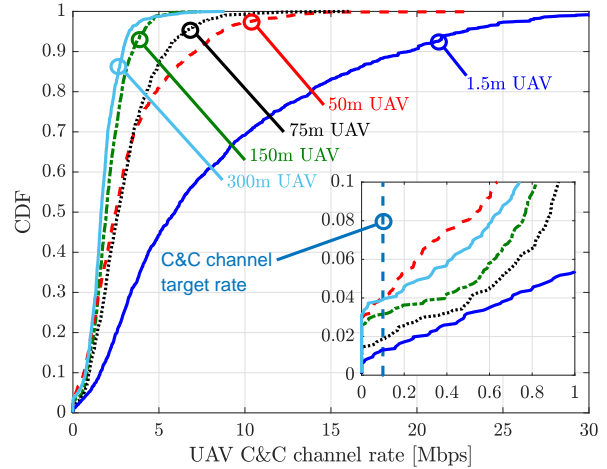}
\caption{CDF of the UAV C\&C channel rates as a function of the UAV height in a multi-user scenario with perfect CSI (Case 3). In the enlargement, the target rate of 100~kbps is also shown as a benchmark.}
\label{fig:MU-McsRate_CDF}
\end{figure}

Similarly to Fig.~\ref{fig:SU-McsRate_CDF} for the single-user case, Fig.~\ref{fig:MU-McsRate_CDF} shows the C\&C rates experienced by a UAV as a function of its height in a multi-user massive MIMO setup. Fig.~\ref{fig:MU-McsRate_CDF} considers the 3GPP Case 3, i.e., one UAV and 14 GUEs per sector. In order to evaluate the gains theoretically achievable with multi-user massive MIMO, in Fig.~\ref{fig:MU-McsRate_CDF} perfect channel state information (CSI) is assumed available at the BSs, i.e., no pilot contamination is accounted for. A realistic channel estimation through SRSs as in (\ref{eqn:PC}) and its effect on the performance of both UAVs and GUEs will be thoroughly discussed in Sec.~\ref{sec:interplay}.

Comparing Fig.~\ref{fig:MU-McsRate_CDF} to Fig.~\ref{fig:SU-McsRate_CDF} provides the reader with the following insights. Unlike single-user cellular networks, massive MIMO networks have the potential to support a 100 kbps UAV C\&C channel with good reliability. Indeed, the data rates in the latter setting are largely improved for three reasons:  \emph{(i)} UAVs enjoy beamforming gain from the serving BS; \emph{(ii)} since most users are GUEs, neighboring BSs tend to point most of their beams downwards, greatly reducing the interference generated at the UAVs; \emph{(iii)} a spatial multiplexing gain is provided by the fact that eight users between UAVs and GUEs are simultaneously allocated to the same PRB. In particular, for all UAV heights under consideration, a massive MIMO network provides UAVs with 100 kbps data rate on the C\&C channel in at least 96\% of the cases.

\section{Interplay between Aerial and Ground Users}
\label{sec:interplay}

\begin{figure}[!t]
\centering
\includegraphics[width=\columnwidth]{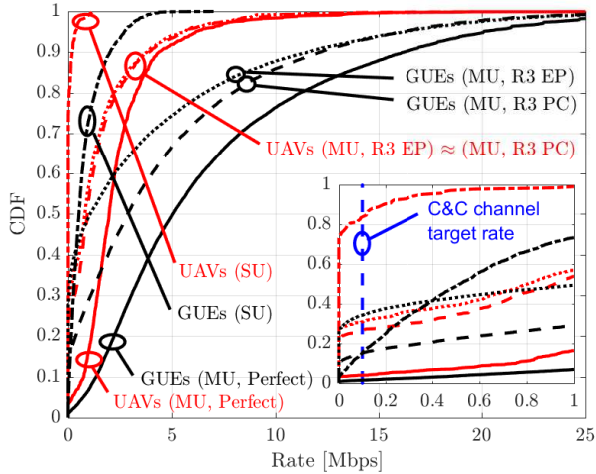}
\caption{Rates in multi-user scenarios under: \emph{(i)} perfect CSI -- ``MU, Perfect'' (solid), \emph{(ii)} SRS Reuse 3 and UL fractional power control -- ``MU, R3 PC'' (dashed), and \emph{(iii)} SRS Reuse 3 with equal power -- ``MU, R3 EP'' (dotted).
The figure also shows the rates in a single-user scenario -- ``SU'' (dash-dot) and, in the enlargement, the UAV C\&C target rate of 100~kbps.}
\label{fig:MU-McsRate_CDF_CSI}
\end{figure}

We now study how supporting the UAV C\&C channel through cellular networks may affect the performance of GUEs. In particular, we thoroughly evaluate the impact of UAVs in both single-user and multi-user settings. Similarly to Fig.~\ref{fig:MU-McsRate_CDF}, Fig.~\ref{fig:MU-McsRate_CDF_CSI} represents the DL rate performance achievable by UAVs and GUEs for the 3GPP Case 3. Not only this figure illustrates the gains provided by multi-user massive MIMO networks, but it also highlights the crucial role played by UL power control through a comparison of three scenarios: \emph{(i)} perfect CSI (``Perfect''), \emph{(ii)} imperfect CSI obtained through SRS Reuse 3 and fractional UL power control (``R3 PC''), and \emph{(iii)} imperfect CSI obtained through SRS Reuse 3 and equal UL power allocation (``R3 EP''). Fig.~\ref{fig:MU-McsRate_CDF_CSI} allows us to conclude this section with the following key takeaways:

\begin{itemize}
\item In spite of having imperfect CSI available at the BSs, the UAV rates greatly improve when moving from a single-user to a multi-user scenario. This is due to the spatial multiplexing gains and the SINR enhancements caused by two complementary effects: the increased beamforming gains and the reduced interference from neighboring BSs, which focus most of their energy downwards.
\item
Pilot contamination can significantly degrade the rate of both UAVs and GUEs when served by massive MIMO BSs. Indeed, the median UAV rates attained under pilot reuse and contamination are reduced to 40\% of those achievable with perfect CSI. In order to close the performance gap caused by pilot contamination, one may resort to more advanced channel estimation and precoding techniques based on multi-cell processing \cite{BjoHoySanICC2017, 6415397}. These schemes leverage in fact channel directionality, which invariably occurs in BS-to-UAV links.
\item 
UL fractional power control does not significantly help to protect the UAV C\&C channel. This is because UAVs generally have favorable propagation conditions with a large number of BSs, which entails that they are the main source of pilot contamination. Instead, UL power control proves a tremendously helpful technique for GUEs severely affected by pilot contamination, as they benefit from the large power reduction of the UAV-generated SRSs against their more conservative power adjustment.
\item 
In networks with heterogeneous ground-and-aerial user populations, massive MIMO also boosts the GUEs data rates. For the cases with imperfect CSI, this is mainly owed to multiplexing gain rather than to SINR gain, since the latter is limited in this scenario due to the severe pilot contamination incurred by GUEs. Indeed, Fig.~\ref{fig:MU-McsRate_CDF_CSI} shows a substantial performance gap between the setups with perfect CSI acquisition and with pilot reuse among sectors. This is because each GUE's SRS is likely to collide with the SRS of at least one UAV in a neighboring cell, with said UAV being likely to experience a strong LoS link with the GUE's serving BS. As for the UAV C\&C channel, massive MIMO is a key enabler, achieving the target rate of 100~kbps in 74\% of the cases even under pilot contamination (``MU, R3 PC'').
\item 
Availing of massive MIMO with perfect CSI would allow to meet said C\&C channel target rate in 96\% of the cases, as opposed to a mere 16\% under single-user setups.
\end{itemize}

\section{Conclusions}
\label{sec:conclusions}

We studied cellular support for the UAV C\&C channel. Under realistic 3GPP assumptions, we compared the performance of a traditional network operating in single-user mode to the one of a multi-user massive MIMO system. We found that the latter can dramatically increase the C\&C channel reliability, meeting the target rate of 100 kbps for UAV heights up to 300\,m in 74\% and 96\% of the cases with and without pilot reuse, respectively. This is owed to a stronger carrier signal, mitigated interference, and a spatial multiplexing gain.

While massive MIMO provides substantial improvements to UAV cellular communications, an additional boost can be pursued with complementary techniques leveraging the time (t), frequency (f), power (p), and spatial (s) domains. These include: protecting neighboring UAVs through almost blank subframes (t/f) \cite{R11717287,R11718267}; alleviating pilot contamination through height-dependent power control (p) and pilot reuse (t/f/s) policies \cite{GalCamLopWCNC2017}; and steering interference towards the channel nullspace of neighboring UAVs (s) \cite{YanGerQueTSP2016,GerGarLop2016,GarGerGal2017}.
\ifCLASSOPTIONcaptionsoff
  \newpage
\fi
\bibliographystyle{IEEEtran}
\bibliography{Strings_Gio,Bib_Gio}
\end{document}